# Extreme Sensitivity of Superconductivity to Stoichiometry in $Fe_{1+\delta}Se$

T. M. McQueen[1], Q. Huang[2], V. Ksenofontov[3], C. Felser[3], Q. Xu[4], H. Zandbergen[4], Y. S. Hor[1], J. Allred[1], A. J. Williams[1], D. Qu[5], J. Checkelsky[5], N. P. Ong[5] and R. J. Cava[1]

[1] Department of Chemistry, Princeton University, Princeton NJ 08544

[2] NIST Center for Neutron Research, National Institute of Standards and Technology, Gaithersburg MD 20899

[3]Institut für Anorganische Chemie und Analytische Chemie, Johannes Gutenberg-Universität, Staudinger Weg 9, D-55099 Mainz, Germany

[4]Department of Nanoscience, TU Delft, The Netherlands

[5]Department of Physics, Princeton University, Princeton NJ 08544

**ABSTRACT**

The recently discovered iron arsenide superconductors appear to display a universal set of characteristic features, including proximity to a magnetically ordered state and robustness of the superconductivity in the presence of disorder. Here we show that superconductivity in $Fe_{1+\delta}Se$, which can be considered the parent compound of the superconducting arsenide family, is destroyed by very small changes in stoichiometry. Further, we show that non-superconducting $Fe_{1+\delta}Se$ is not magnetically ordered down to 5 *K*. These results suggest that robust superconductivity and immediate instability against an ordered magnetic state should not be considered as intrinsic characteristics of iron-based superconducting systems.

INTRODUCTION

Superconductivity was discovered in 2008 in LaFeAsO$_{1-x}$F$_x$[1] with a T$_c$ of 26 *K*. The superconductivity in this arsenide, and the previously reported LaFePO$_{1-x}$F$_x$[2], is unexpected because most Fe-based compounds display magnetic ordering at low temperatures. This new family of superconductors, based on Fe$_2$X$_2$ (X = P,As) layers of edge-sharing FeX$_4$ tetrahedra, has expanded to include doped but oxygen-free systems, including K$^+$/Na$^+$-doped AFe$_2$As$_2$ (A = Ba,Sr,Ca)[3,4], and (Li,Na)$_x$FeAs[5,6]. Recently, superconductivity at 8 *K* has been reported[7] in chemically analogous FeSe in its tetragonal form (β-FeSe; recent publications have referred to this, improperly, as the α form. In phase diagrams and the original literature, it is the β form that is tetragonal (cf refs 8, 9), although a few, e.g. ref. 10, refer to tetragonal FeSe as the α form. α is used here to designate the stoichiometric NiAs-type variant.), and the superconductivity is reported to increase to 27 *K* under modest pressure[11]. The Fe$_2$Se$_2$ layers in β-FeSe (inset Fig. 1) are analogous to the Fe$_2$As$_2$ and Fe$_2$P$_2$ layers in the pnictide and oxypnictide superconductors. The initial report attributed the superconductivity to a highly selenium deficient phase, FeSe$_{0.82}$ (Fe$_{1.22}$Se)[7]. This was quickly followed by a combined x-ray and neutron diffraction study that arrived at a composition of FeSe$_{0.92(1)}$ (Fe$_{1.09}$Se)[12] for the superconductor. Both of these formulas fall well outside the narrow composition range, Fe$_{1.01}$Se-Fe$_{1.04}$Se, reported for β-FeSe more than thirty years ago[13,14]. Here we show that, when prepared so as to prevent the formation of spurious oxides and oxygen defects in the phase, superconducting β-FeSe is much closer to stoichiometric than the recent reports indicate. Further, we find that the superconducting transition temperature is critically dependent on extremely small changes in the iron stoichiometry. The highest transition temperatures, T$_c$ ~ 8.5 *K*, are found when the compound is closest to stoichiometric, with formula β-Fe$_{1.01}$Se. With a little more iron excess, at composition

β-Fe$_{1.02}$Se, T$_c$ drops to 5 *K*, and, with slightly more iron, β-Fe$_{1.03}$Se is non-superconducting down to 0.6 *K*. Non-superconducting β-Fe$_{1.03}$Se does not exhibit a long range ordered magnetic state, but only the suggestion of spin fluctuations at low temperature. Subtle differences in the structure indicate that there may be a difference in defect chemistry between superconducting and non-superconducting compositions. Our results indicate that superconductivity in β-FeSe is only borderline stable, and that it does not directly compete with a magnetically ordered state.

**EXPERIMENTAL**

Except for "Fe$_{1.06}$Se" (see below), all samples were prepared from iron pieces (Johnson-Matthey, 99.98%) and selenium shot (Alfa-Aesar, 99.999%). Stoichiometric quantities of freshly polished iron and selenium shot were loaded into cleaned and dried silica tubes, and sealed under vacuum with a piece of cleaned carbon inside (but not in physical contact with the sample). These tubes were sealed in a second evacuated silica ampoule and placed in a furnace at 750 *°C*. The temperature was held constant until the Se vapor had disappeared (3-5 days), and then increased to 1075 *°C* for three days, followed by a fast decrease to 420 *°C*. This temperature was held for two days before the tubes were quenched in -13 *°C* brine. Small pieces were then loaded into small silica ampoules and annealed at various temperatures (300-500 *°C*) for two days followed by quenching in -13 *°C* brine. Fast quenching was required for reproducible behavior. All samples are stable for short periods of time in air, but were protected from oxidation in air by storage in an argon glove box. $^{57}$Fe Mössbauer spectra were recorded in a transmission geometry using a conventional constant-acceleration spectrometer and a helium bath cryostat. The Recoil Mössbauer Analysis Software was used to fit the experimental spectra. Isomer shift values are quoted relative to α-Fe at 293 *K*. DC magnetization measurements were performed on a quantum design physical property measurement system (QD-PPMS) using powdered samples to minimize demagnetization effects. Based on low field M(H) curves at 2 $K$, the absolute error in the dc magnetization values is estimated to be less than 10%. High resolution neutron powder diffraction (NPD) data were collected using the BT-1 high-resolution powder diffractometer at the NIST Center for Neutron Research, employing a Cu (311) monochromator to produce a monochromatic neutron beam of wavelength 1.5403 Å. Collimators with horizontal divergences of 15′, 20′, and 7′ full width at half maximum were used before and after the monochromator, and after the sample, respectively. The intensities were measured in steps of 0.05° in the 2θ range 3-168°. The structure analysis was performed using the program GSAS with EXPGUI[15, 16]. The neutron scattering amplitudes used in the refinements were 0.945, and 0.797 ($\times 10^{-12}$ $cm$) for Fe and Se, respectively. Specific heat measurements were done on polycrystalline pellets in a QD-PPMS equipped with a $^3$He refrigerator. Resistivity measurements were done in an Oxford cryostat using the four probe method, at a frequency of 13 Hz and a current of 0.1 mA. Thermopower measurements were done using a custom-built helium probe-head (a MMR sample stage reduced in size to fit in the cryogenic probe) and MMR technologies electronics. The double reference measurement technique was used, with constantan wire as the reference. X-ray powder diffraction (XRD) was done using a Bruker D8-Focus employing Cu-Kα radiation with a diffracted beam monochromator.

**RESULTS AND DISCUSSION**

Our initial attempts to prepare phase-pure β-FeSe employed the methods recently described[7, 12], starting with 'freshly cleaned' Fe (Alfa-Aesar, 99.95%) in powder form. In agreement with those reports, we found that a significant iron excess, in our case $Fe_{1.06}Se$, was needed to make a sample that appeared to be "phase pure" by laboratory x-ray diffraction (XRD). When prepared from very clean starting materials, however, specifically taking care to exclude oxygen (see above), we found that the composition needed to yield a single phase specimen was close to $Fe_{1.01}Se$. The origin of this discrepancy was studied using several techniques. Fig. 2(a) shows, in the left inset, a region of the neutron powder diffraction (NPD) patterns for samples of $Fe_{1.06}Se$ and $Fe_{1.01}Se$, revealing that the sample prepared from Fe powder ("$Fe_{1.06}Se$") is contaminated with $Fe_3O_4$[17], observed because NPD is more sensitive than laboratory XRD to the presence of impurities. The presence of iron oxide explains why excess iron is needed to obtain a "pure" specimen under normal synthetic conditions. Furthermore, energy-dispersive x-ray spectroscopy (EDX) measurements in the transmission electron microscope on crystallites of the β-FeSe phase from $Fe_{1.06}Se$ (Fig. 2(a), right inset) showed substantial oxygen contamination. Electron energy loss spectroscopy (EELS) of the iron $L_3$ and $L_2$ edges (Fig. 2(a)) confirm that there is significant oxidation of the iron in $Fe_{1.06}Se$, i.e. oxygen is bound to the iron atoms within the β-FeSe such that the formula is $Fe_{1+\delta}SeO_y$. EDX and EELS on a $Fe_{1.01}Se$ sample ($Fe_{1.01}Se$ annealed at 300°C) made from very clean starting materials shows no oxygen by EDX and no unexpected oxidation of the iron by EELS (Fig. 2(a)).

The composition of the β-FeSe phase was confirmed to be nearly stoichiometric by Rietveld refinements of NPD data on both "$Fe_{1.06}Se$" and $Fe_{1.01}Se$ annealed at 300°C. When freely refined, the composition of the tetragonal phase in "$Fe_{1.06}Se$" is nearly stoichiometric (Table I, col. 1), and a similar result was obtained for $Fe_{1.01}Se$ (Table I, col. 2 and Fig. 1). To best determine the stoichiometry and to see if we could locate the origin of any non-stoichiometry, we performed free fits to the data as a function of fixed doping levels, with excess iron in interstitial sites[18] and with selenium vacancies. The refinement agreement statistics $R_{wp}$ (minimum for the best agreement) for those refinements as a function of hypothetical stoichiometry are plotted in the inset to Fig. 1. The best agreement is centered at the stoichiometric FeSe composition, with the breadth of the minimum indicating a composition of $Fe_{1.01\pm0.02}Se$. Thus, although these measurements do not have sufficient sensitivity to determine the stoichiometry to better than ±0.02, the nearly ideal stoichiometry of the β-FeSe phase, in agreement with the results in the older literature,[13, 14] is clearly confirmed. The compositions $Fe_{1.09}Se$[12] and $Fe_{1.22}Se$[7] are not consistent with these data, as the refinement statistics are markedly worse (near the top right of the right inset, Fig. 1 for $Fe_{1.09}Se$ with Se vacancies, and off scale for $Fe_{1.22}Se$).

The magnetic characterization of the superconducting transition in a selection of our samples, measured by low field dc magnetization, is shown in Fig. 2(b). The data show that $Fe_{1.01}Se$ prepared at a temperature of 300 °C is a superconductor with a sharp transition near ~8.5 K. This is in contrast to a sample poisoned with oxygen (dashed line, similar to the original literature report[7]). Furthermore, a sample that is slightly more iron rich, $Fe_{1.02}Se$ annealed at 380 °C, shows a reduced $T_c$. Finally superconductivity is absent for $Fe_{1.03}Se$ annealed at 400 °C. This reflects an extreme dependence of the superconducting properties on preparative conditions, including stoichiometry and temperature.

Our data indicate that the superconductivity, the stoichiometry, and the crystal structure are correlated in the β-FeSe-type phase. Fig. 3(a) shows the superconducting transition temperature $T_c$ versus the crystallographic c/a ratio for a number of the samples in this system. Several features are evident. Samples prepared at lower temperatures or with lower iron content display c/a ratios just above 1.464 and also display the highest transition temperatures. Higher iron contents or higher synthesis temperatures yield larger c/a ratios and reduced $T_c$'s. Samples prepared with the highest iron content, $Fe_{1.03}Se$, show no superconductivity to 0.6 K, irrespective of the synthesis temperature. These samples also display abnormally small c/a ratios, near 1.461. Two distinct structure/superconductivity regions are therefore clearly seen in Fig. 3(a). The inset of Fig. 3(a) shows the dependence of the c/a ratio on starting composition for two representative preparation temperatures. In both cases, c/a initially rises with increasing Fe content, but by the composition $Fe_{1.03}Se$ the c/a ratio is much reduced.

We postulate that this peculiar dependence of c/a ratio on iron content and the accompanying dramatic disappearance of superconductivity for $Fe_{1.03}Se$ are due to a change in how the non-stoichiometry is accommodated in the phase. The most likely scenario is a change from Se vacancies at low Fe excess to Fe interstitials at high Fe excess. This may also explain why the c/a ratio changes for a given nominal composition as the synthesis temperature is changed - the stability of the different types of defects is no doubt temperature dependent. Further studies will be of interest to elucidate the origin of this behavior.

On the basis of these experiments, we construct a phase diagram for the Fe-Se system near the 1:1 stoichiometry in Fig. 3(b). Samples quenched from above 455 °C contain significant fractions of three-phases (not possible for equilibrium conditions in a binary system). This is consistent with the proposal in the old literature that iron-rich, hexagonal δ-FeSe, stable at high temperatures, converts to tetragonal FeSe on cooling[8]. Thus we assign 455 °C as the upper limit of temperature stability for β-FeSe. This agrees well with the reported decomposition temperature of 457 °C[9]. Additionally, we find that β-FeSe is unstable at low temperatures: there is a slow conversion of the tetragonal β-$Fe_{1+\delta}Se$ phase to a hexagonal NiAs structure type (α-FeSe) phase, with larger lattice parameters than are found for "$Fe_7Se_8$"[19], below approximately 300 °C. This hexagonal phase is non-superconducting down to 0.6 K. Since the best superconducting properties of the β-$Fe_{1+\delta}Se$ phase appear with the lowest iron contents at the lowest synthesis temperatures, this conversion to the NiAs form at low temperatures ultimately puts a limit on the maximum $T_c$ obtainable in this system.

Further evidence of the extreme dependence of the properties of β-$Fe_{1+\delta}Se$ on stoichiometry and preparation conditions can be seen in the low temperature specific heats, which are shown for four compositions, $Fe_{1.01}Se$-300°C, $Fe_{1.01}Se$-330°C, $Fe_{1.02}Se$-380°C, and $Fe_{1.03}Se$-400°C, in Fig. 4. The raw data clearly show the presence of excess specific heat associated with the superconducting transition, and that $T_c$ moves to lower temperatures with increasing iron excess. Quantitative analysis of the electronic and magnetic contributions to the specific heat requires the removal of the lattice contribution, which can't be done in the usual fashion in this system because no portions of the C/T vs. $T^2$ plots are linear, implying that the lattice contribution is not simply given by $\beta_3 T^3$ up to 15 $K$. As such, we fit the 10-15 $K$ region of $Fe_{1.01}Se$-300°C to $C = \gamma T + \beta_3 T^3 + \beta_5 T^5$, where the first term accounts for the normal-state electronic contribution, and the second and third terms account for the lattice contribution. Parameters are given in Table II. The Debye temperature calculated from $\beta_3$ is $\theta_D = 200 K$. (This explains why $\beta_3 T^3$ is not sufficient to account for the lattice contribution, as $\beta_3 T^3$ is generally only good up to $\frac{\theta_D}{50} = 4K$ [20].) Subtracting the lattice contribution with the fitted $\beta_3$ and $\beta_5$ values gives the residual electronic contribution, shown in the inset to Fig. 4. The normal-state Sommerfeld coefficient is then estimated as $\gamma = 5.4(3) mJmol^{-1}K^{-2}$. A very well defined, sharp transition to the superconducting state is seen. From this data, using the equal entropy construction, we estimate that the normalized specific heat jump at $T_c$ is $\Delta C/\gamma T_c = 1.3(1)$, which is in good agreement with the BCS expected value of 1.4. This confirms the bulk nature of the superconductivity below 8.5 $K$ in $Fe_{1.01}Se$-300°C. The amount of excess entropy lost near $T_c$ is well balanced by the entropy difference between the normal and superconducting states at low temperature, therefore supporting the validity of the lattice subtraction.

Surprisingly, the data show (inset, Fig. 4) that there is a second specific heat anomaly at 1 $K$ in the optimal superconducting sample. To characterize the dependence of this anomaly on the stoichiometry, specific heat data on three other samples is also shown. A fit of the 10 – 15 $K$ region of the specific heat of $Fe_{1.03}Se$ to $C = \gamma T + \beta_3 T^3 + \beta_5 T^5$ gives parameters that are similar to those obtained for $Fe_{1.01}Se$-300°C (Table II). The origin of the differences is unclear; the lower Sommerfeld coefficient may reflect a change in the electronic state of $Fe_{1.03}Se$. The differences in $\beta_3$ and $\beta_5$ may indicate that there are extra contributions to the specific heat (e.g. spin fluctuations). Since the data on the intermediate samples does not extend to sufficiently high temperatures (15 $K$) to permit separate fits of the high temperature region to remove the lattice contributions, we employed the approximation that the lattice contribution to the specific heat (the $\beta_3$ and $\beta_5$ parameters) for superconducting $Fe_{1.02}Se$-380°C and $Fe_{1.01}Se$-330°C are the same as for $Fe_{1.01}Se$-300°C, and employ the as-fit parameters for $Fe_{1.03}Se$ (The qualitative features mentioned below do not change when the same $\beta_3$ and $\beta_5$ terms are used for all samples.), and present the results in the inset to Fig. 4. As Fe content increases to x = 1.02, the superconducting anomaly shifts to lower temperature and decreases in magnitude. Simultaneously, the 1 $K$ anomaly increases dramatically. In $Fe_{1.03}Se$-400°C, the 1 $K$ anomaly is not present and a third kind of behavior is observed – a slowly rising specific heat with decreasing temperature. This contribution is also likely present in the $Fe_{1.02}Se$ sample. The low temperature upturns in $Fe_{1.02}Se$-380°C and $Fe_{1.03}Se$-400°C are qualitatively consistent with spin fluctuations, but may also be attributable to lattice defects, small amounts of impurity phases, or some type of very low temperature magnetic ordering. The electronic contribution to the specific heat seems to decrease as Fe content is increased. Further studies are needed to determine the origin of these low temperature specific heat anomalies, and to confirm the change in the electronic contribution.

Temperature-dependent resistivity measurements on $Fe_{1.01}Se$-300°C and $Fe_{1.03}Se$-400°C show differences between the superconducting and non-superconducting stoichiometries of β-$Fe_{1+\delta}Se$ (Fig. 5). $Fe_{1.01}Se$-300°C displays metallic resistivity, with a residual resistivity ratio (RRR) of 10, reasonable for a measurement on a polycrystalline, metallic sample. It also shows a superconducting transition at 9 $K$, consistent with the susceptibility and specific heat measurements. Furthermore, there is a kink near 90 $K$ (see Fig. 5 inset), corresponding to the temperature of the previously reported structural transition[12]. In contrast, $Fe_{1.03}Se$-400°C shows a broad feature in the resistivity around 90 $K$ and no superconductivity. Furthermore, the magnitude of the resistivity at room temperature is higher, and the RRR (=2) is reduced, when compared to $Fe_{1.01}Se$-300°C. These observations are consistent with the presence of a larger number of defects in the higher Fe content phase. Despite the substantial effect of stoichiometry on the resistivity, the Seebeck coefficients (α) are qualitatively similar for superconducting and non-superconducting β-$Fe_{1+\delta}Se$ (Fig. 5 inset). In both cases, α is small and positive at room temperature, changes sign near 230 $K$, and goes through a broad (negative) maximum near the structural phase transition around 90 $K$. Like in the resistivity, the transition around 90 $K$ appears broadened in $Fe_{1.03}Se$, but they are otherwise very similar. The change of sign implies that electrons and holes contribute nearly equally to the conduction. Additionally, the broadening of the kink near 90 $K$ in both datasets suggests that the change in defects on going from β-$Fe_{1.01}Se$ to β-$Fe_{1.03}Se$ may be having an impact on the structural phase transition.

To determine whether the state competing with superconductivity in β-FeSe has a magnetic origin, we employ Mössbauer spectroscopy as a sensitive local probe for the presence of magnetism at the iron sites. Representative spectra are shown in Fig. 6. Despite the large differences in the superconducting properties, the Mössbauer spectra for all the oxygen-free samples[21] are very similar. A single quadrupole paramagnetic doublet is sufficient to describe all the spectra. The hyperfine parameters (Table III) agree well with those previously found.[10] The quadrupole splitting can be attributed to the distortion from tetragonal symmetry of the local surrounding of iron atoms. The isomer shift and quadrupole splitting are both increased at 80 $K$ (below the symmetry-lowering phase transition[12]), but are then essentially unchanged at 5 $K$, whether the sample is superconducting or not. The linewidths do increase slightly on cooling ($\Gamma$ = 0.15(1) mm/s at 295 $K$, $\Gamma$ = 0.19(1) mm/s at 5 $K$), but this is expected. More importantly, although the linewidth of the doublet in β-Fe$_{1.03}$Se may be marginally larger than that in β-Fe$_{1.01}$Se at 5 $K$ (Fig. 5), the spectrum does not display the additional dramatic splitting (into a sextet) expected for an ordered magnetic phase. This is in sharp contrast to undoped LaOFeAs, for example, which shows a clear splitting of the Mössbauer spectrum into a sextet below the spin density wave (SDW) transition[22]. This shows that the electronic state in β-Fe$_{1.03}$Se is not magnetically ordered in nature. The presence of magnetic fluctuations on a timescale shorter than the Mössbauer timescale ($10^{-7}$ s) cannot be ruled out, but there is no long range magnetic ordering at 5 $K$.

**CONCLUSION**

Our results indicate that the superconductivity in β-FeSe is very sensitive to composition and disorder even though many of the basic characteristics of the superconducting and non-superconducting compositions are quite similar. That a small number of defects is important is surprising because the high upper critical field (800 $kOe^7$) and chemical similarity to the FeAs-based superconductors implies that superconductivity in β-FeSe should be more robust. This sensitivity to defects likely extends to other members of this family, and may explain the conflicting reports about superconductivity in stoichiometric LaFePO[2, 23-26]. Furthermore, the fact that we do not observe magnetic ordering down to 5 $K$ in non-superconducting β-Fe$_{1.03}$Se implies either that β-FeSe is fundamentally different from the FeAs-based compounds, or that superconductivity does not directly arise from a competing, ordered magnetic state in all members of this superconducting family (spin correlations are not ruled out). The former seems unlikely, as density functional theory calculations on FeSe[27] show the same general features as in the FeAs systems – namely, a highly two dimensional Fermi surface and propensity for SDW behavior. If the latter is the case, it then implies that magnetically ordered and superconducting states are not as transparently related in this family as they currently appear. It may be that further doping (beyond the limits of the binary phase diagram) will eventually induce a SDW state in β-FeSe, and that β-Fe$_{1.03}$Se is in an intermediate state such as the pseudogap state in the cuprates or the quantum critical state in other systems. As such, these results suggest that understanding the electronic state of β-Fe$_{1.03}$Se will be critical in understanding the superconductivity in the iron-based systems as a whole.

**ACKNOWLEDGEMENTS**

The work at Princeton was supported primarily by the US Department of Energy, Division of Basic Energy Sciences, Grant DE-FG02-98ER45706, and, in part, by the NSF MRSEC program, grant DMR-0819860. T. M. McQueen gratefully acknowledges support of the National Science Foundation Graduate Research Fellowship program.

Table I. Refined structural parameters for two samples of β-FeSe at 298 $K$ from powder neutron data. Space group *P4/nmm* (#129). Atomic positions: **Fe**: 2a (3/4,1/4,0), **Se**: 2c (1/4,1/4,z). Lattice parameters are in units of Å, and thermal parameters are in units of $10^{-2}$ Å$^2$. "Fe$_{1.06}$Se" contains small secondary phases of Fe and Fe$_3$O$_4$. The β-Fe$_{1.01}$Se sample employed contains very small amounts of Fe, Fe$_7$Se$_8$ and α-FeSe.

|    |          | "Fe$_{1.06}$Se" | Fe$_{1.01}$Se |
|----|----------|-----------------|---------------|
|    | a        | 3.7747(1)       | 3.7734(1)     |
|    | c        | 5.5229(1)       | 5.5258(1)     |
| **Fe** | $U_{11}$ | 0.87(2)     | 0.63(3)       |
|    | $U_{33}$ | 2.02(4)         | 2.41(5)       |
|    | occ      | 0.987(6)        | 0.997(3)      |
| **Se** | $U_{iso}$ | 1.35(3)    | 1.31(3)       |
|    | z        | 0.2669(2)       | 0.2672(1)     |
|    | $\chi^2$ | 1.727           | 2.117         |
|    | $R_{wp}$ | 6.42%           | 6.56%         |
|    | $R_p$    | 5.15%           | 5.30%         |
|    | $R(F^2)$ | 6.04%           | 7.42%         |

**Table II**. Values obtained from fits of the 10-15 $K$ regions of the heat capacity of $Fe_{1.01}Se$-300°C and $Fe_{1.03}Se$ to $C = \gamma T + \beta_3 T^3 + \beta_5 T^5$ (see text).

| | $\gamma$ (mJ mol$^{-1}$ K$^{-2}$) | $\beta_3$ (mJ mol$^{-1}$ K$^{-4}$) | $\beta_5$ (mJ mol$^{-1}$ K$^{-6}$) |
|---|---|---|---|
| $Fe_{1.01}Se$-300°C | 5.4(3) | 0.463(5) | -2.8(2)·10$^{-4}$ |
| $Fe_{1.03}Se$ | 1.3(6) | 0.496(8) | -4.2(2)·10$^{-4}$ |

**Table III**. Mössbauer isomer shift and quadrupole splitting values for select β-Fe$_{1+\delta}$Se samples at various temperatures.

| T= | | β-Fe$_{1.01}$Se | β-Fe$_{1.03}$Se |
|---|---|---|---|
| **295 K** | δ (mm/s) | 0.46(1) | 0.47(1) |
| | ΔE$_Q$ (mm/s) | 0.25(2) | 0.26(1) |
| **80 K** | δ (mm/s) | 0.57(1) | 0.55(2) |
| | ΔE$_Q$ (mm/s) | 0.29(2) | 0.30(2) |
| **5 K** | δ (mm/s) | 0.57(1) | 0.59(3) |
| | ΔE$_Q$ (mm/s) | 0.30(1) | 0.34(3) |

**Fig 1**. (Color Online) Rietveld refinement of 298 K NPD data of β-Fe$_{1.01}$Se-300°C. The left inset shows the fit statistic R$_{wp}$ plotted versus Fe-interstitals (left) and Se-vacancies (right). From these data it is not possible to determine the origin of the 1% non-stoichiometry, but this shows that the formula of superconducting β-FeSe must be within ~2% of stoichiometric. The right inset shows the structure of β-FeSe.

**Fig. 2**. (Color Online) (a) EELS, NPD (left inset) and EDX (right inset) data on β-Fe$_{1.01}$Se and "Fe$_{1.06}$Se". The EDX analysis on crystallites of the β-Fe$_{1+\delta}$Se phase in the Transmission electron microscope shows the presence of oxygen in "Fe$_{1.06}$Se", and the EELS data of the L$_3$ and L$_2$ peaks confirm that the oxygen is bonded to the iron (arrows). This is in addition to the Fe$_3$O$_4$ present in "Fe$_{1.06}$Se". (b) Low field susceptibility data of various Fe$_x$Se samples, showing that β-Fe$_{1.03}$Se is non-superconducting and that superconductivity improves going from β-Fe$_{1.02}$Se to β-Fe$_{1.01}$Se. For comparison, the susceptibility of a sample poisoned with oxygen, similar to previous work, is also shown (dashed line).

**Fig 3**. (Color Online) (a) Dependence of superconducting temperature (defined as the midpoint of the dc susceptibility transition) on c/a ratio. The inset shows the dependence of c/a ratio on synthesis temperature and nominal composition. (b) Phase diagram derived from the samples shown in (a) and others (not shown). Actual compositions of the samples were estimated from the fraction of impurity phases present (Fe$_7$Se$_8$ and Fe metal) by XRD and/or room temperature M(H) curves. Below 300 °C, β-Fe$_{1+\delta}$Se slowly converts to α-Fe$_x$Se, which has the NiAs structure type and is non-superconducting above 1.8 K. The c/a ratios of β-FeSe also suggest a change in defect type as Fe content increases within the phase (represented by the vertical dotted line and shading).

**Fig. 4**. (Color Online) Low temperature specific heat of β-Fe$_{1.01}$Se-300°C, β-Fe$_{1.01}$Se-330°C, β-Fe$_{1.02}$Se-380°C, and β-Fe$_{1.03}$Se-400°C. The inset shows the data after subtraction of a lattice contribution (see text).

**Fig. 5**. (Color Online) Resistivity data show that β-Fe$_{1.01}$Se is a good metal with a superconducting transition near 9 K whereas β-Fe$_{1.03}$Se is metallic but with a low residual resistivity ratio. β-Fe$_{1.01}$Se shows a change in slope around 90 K, corresponding to the temperature of the previously reported structural distortion[12], but the transition in β-Fe$_{1.03}$Se is broadened (first derivative plotted in top inset). Bottom inset: the Seebeck coefficient of β-Fe$_{1.01}$Se and β-Fe$_{1.03}$Se are similar in magnitude and change sign around 230 K. They also show a change at the structural phase transition, but the transition in β-Fe$_{1.03}$Se is significantly broader. This suggests that defects have a substantial impact on the phase transition.

**Fig 6**. (Color Online) $^{57}$Fe Mössbauer spectra at 295, and 5 K. There are no significant differences between β-Fe$_{1.01}$Se and β-Fe$_{1.03}$Se, despite the fact that β-Fe$_{1.01}$Se is superconducting at 8.5 K and β-Fe$_{1.03}$Se shows no superconductivity above 0.6 K. There is no sign of magnetic ordering in these samples. Extra magnetic contributions to the Mössbauer spectra only appear in samples poisoned with oxygen (data shown in insets).

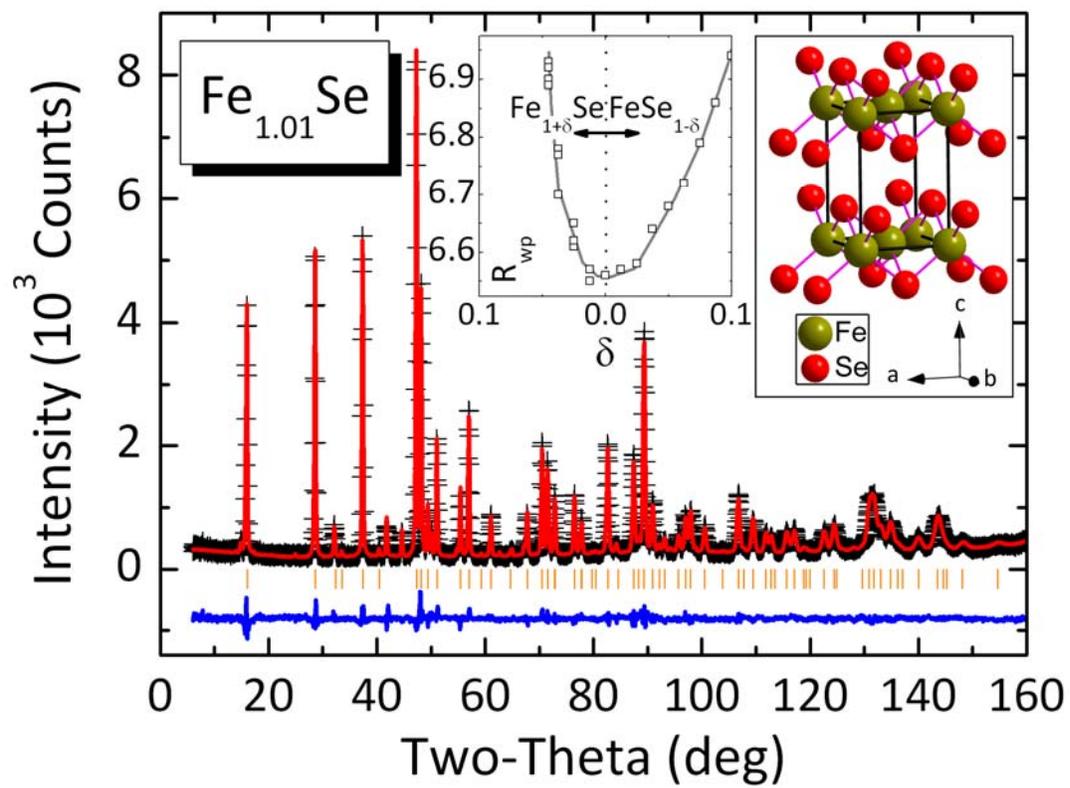

Figure 1.

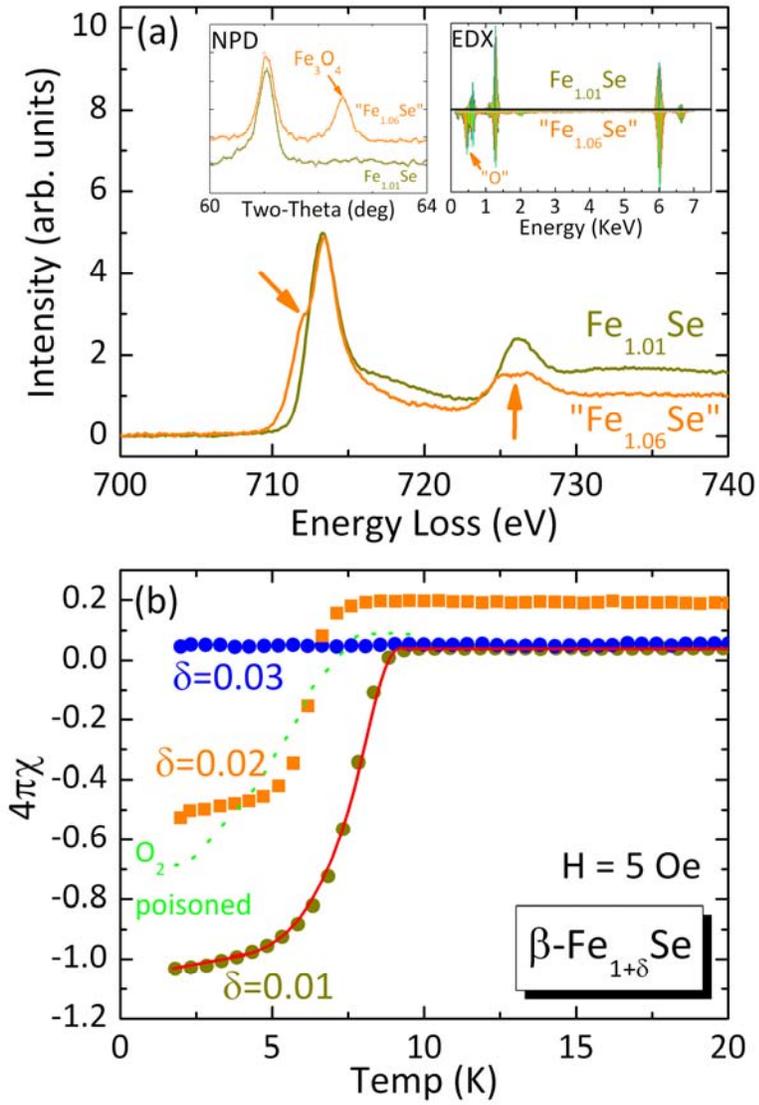

Figure 2.

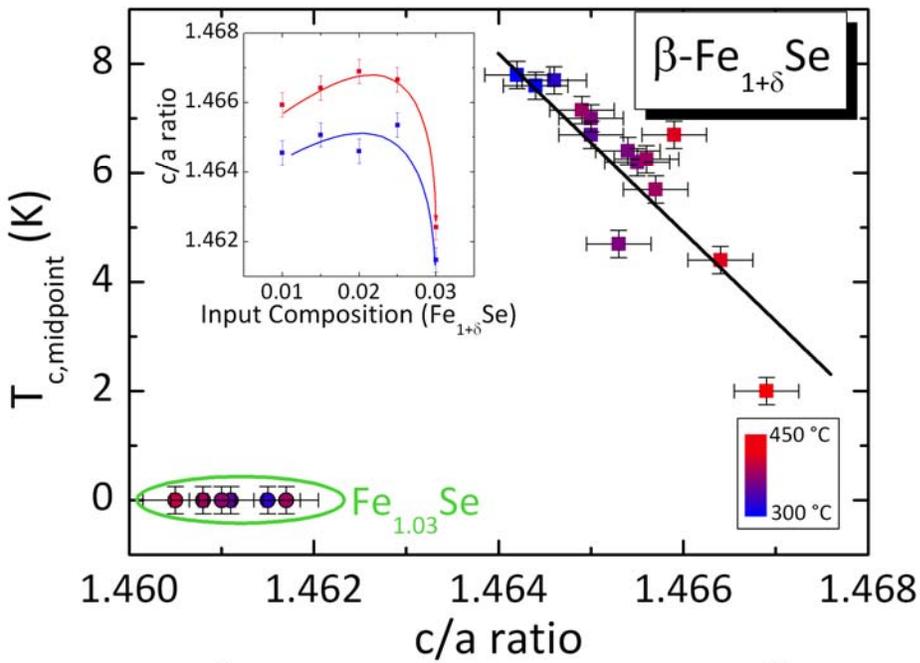
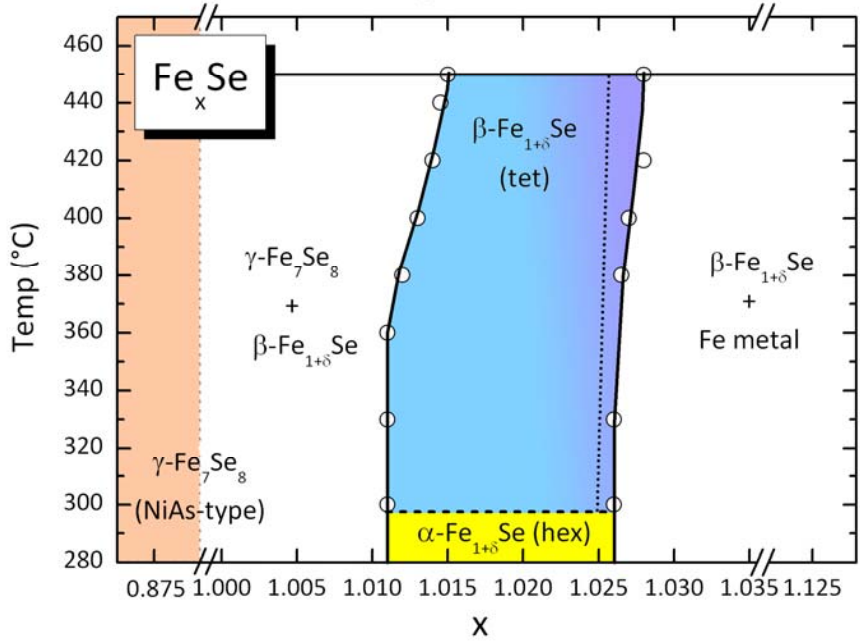

Figure 3.

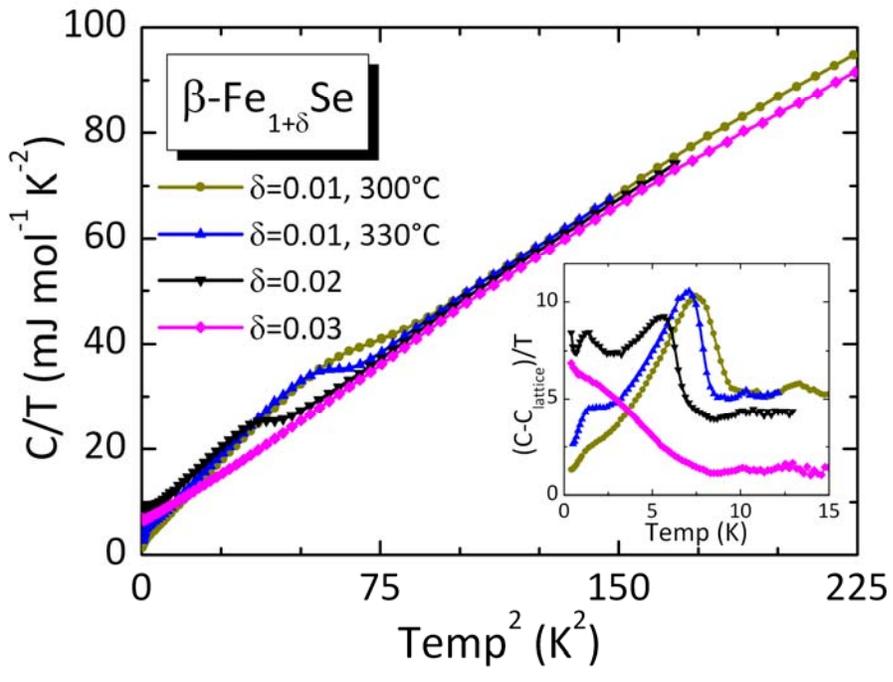

Figure 4.

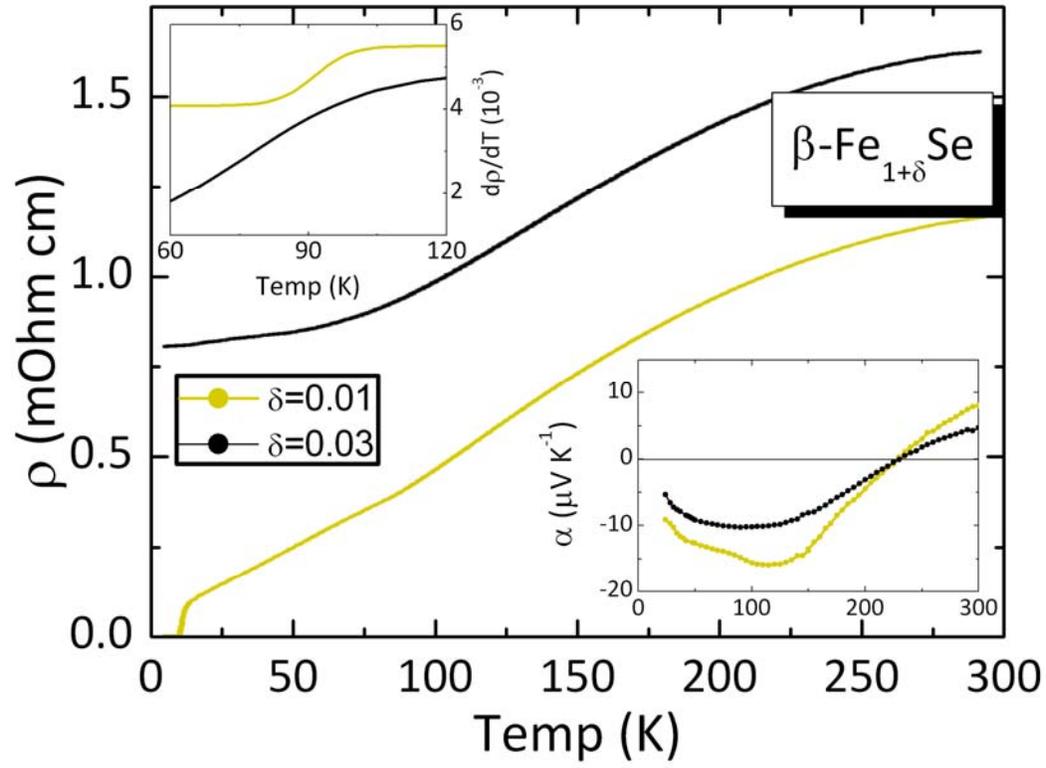

Figure 5.

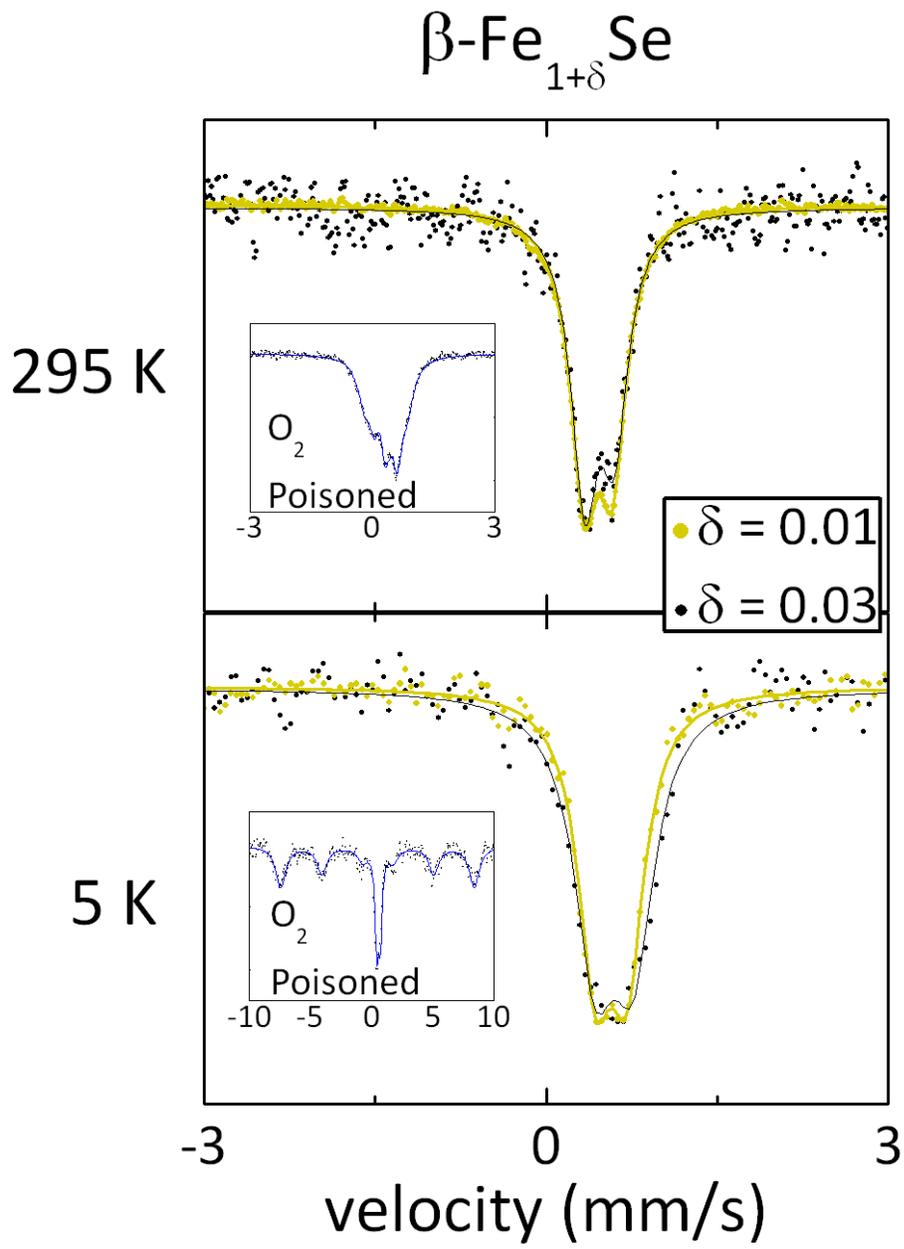

Figure 6.